# Barriers to the Integration of Information Technology within Early Childhood Education and Care Organisations: A Review of the Literature

**Melinda Plumb**
School of Management, Operations and Marketing
University of Wollongong
Wollongong, NSW, Australia
Email: map016@uowmail.edu.au

**Karlheinz Kautz**
School of Management, Operations and Marketing
University of Wollongong
Wollongong, NSW, Australia
Email: kautz@uow.edu.au

## Abstract

Employees of early childhood education and care (ECEC) organisations may experience a wide range of barriers as they attempt to integrate information technology (IT) into their work practices. However, studies within the ECEC organisational literature which attempt to identify and understand these barriers are scant. This literature review is the first to present consolidated findings from the body of knowledge on barriers to the integration of IT within ECEC organisations. In addition to highlighting limitations and gaps in the literature, it proposes a tri-perspective framework to provide for future research to develop a deeper understanding of not only what barriers exist but also how they interrelate and shape the IT integration process and the work practices of ECEC organisational employees.

**Keywords**

Information technology, integration, barriers, early childhood organisations, educators

## 1 Introduction

Information technology (IT) is becoming an increasingly common feature within early childhood education and care (ECEC) organisations, as the once-fierce debate on whether IT use is appropriate in early childhood settings and whether it inhibits or enhances young children's learning and development is less polarised now (Nikolopoulou and Gialamas 2013). Countries around the world such as Australia, New Zealand, the United Kingdom (U.K.), Portugal, Sweden and Denmark are formally recognising the role IT can play within ECEC organisations (Plumb et al. 2013) and are moving to integrate IT in early childhood curricula. However according to Ertmer (1999) educators continue to grapple with both "practical and philosophical problems" posed by the process of attempting to successfully integrate IT into their classrooms, stating that it is "almost guaranteed" that they will experience a wide range of barriers (p.50). Educators who work in ECEC organisations provide education and care services to children from birth to prior to their attendance at school and are not exempt from encountering barriers as they attempt to successfully integrate IT into their work practices. The majority of literature examining IT in educational organisations has focused primarily on educators in schools (Fenty and McKendry Anderson 2014; Nikolopoulou and Gialamas 2013; Wood et al. 2008), and this includes literature that specifically examines barriers to IT integration (Blackwell et al. 2014; Blackwell et al. 2013). The literature on barriers to IT integration by educators in ECEC organisations is limited to a "small number of empirical studies" (Nikolopoulou and Gialamas 2013 p.3) which leaves "much unknown about early childhood educators, who may pose different concerns and face different barriers to technology integration" (Blackwell 2013 p.235).

Although several literature reviews of the barriers to the integration of IT in schools (BECTA 2004; Hew and Brush 2007) and universities (Reid 2014) exist, no literature reviews were found to exist for ECEC organisations. This paper therefore presents the first comprehensive literature review on barriers to the integration of IT within such organisations. It extends the prior findings of Plumb et al. (2013) who have previously reviewed the literature on touch screen IT adoption and utilisation in ECEC organisations, broadening the field to all forms of IT but narrowing the phenomenon of study to the barriers experienced by ECEC organisation employees as they move to integrate IT into their work practices. The remainder of this paper proceeds as follows: first the literature review methodology is presented, outlining the search and filtering techniques utilised in order to obtain the set of papers considered by this literature review; identified barriers from the literature are then descriptively presented in the results section; and finally a discussion of gaps and issues with the literature and a tri-perspective framework for understanding barriers to IT integration is presented, highlighting opportunities for further research.

## 2 Methodology

The assessment of articles for inclusion in this literature review proceeded in three stages, similar to those identified by Wolfswinkel et al. (2013) in their article on grounded theory as a method for reviewing literature.

Stage *one* involved defining the scope of the review, inclusion and exclusion criteria, and search terms. We have chosen to utilise multiple journal databases in order to 'cast our net wide' for our source material. The Scopus database was chosen for its multidisciplinary index and reputation as "the largest abstract and citation database of peer-reviewed literature" (Elsevier B.V. 2015) and the Web of Science™ database as a supplementary multidisciplinary index which also indexes the Association for Information Systems' Senior Scholars 'basket of 8' journals (Association for Information Systems 2015). Due to the nature of the phenomenon under study EdITLib, a topic-specific database, was additionally utilised as it is considered to be "the premiere online resource for aggregated, peer-reviewed research on the latest developments and applications in Educational Technologies" (Global U. 2015). Additionally the Education Research Complete and ERIC (Education Resources Information Centre) databases were utilised via EBSCO. Table 1 below details the keywords selected for use in the database searches.

| Concept | Search terms | Justification |
| --- | --- | --- |
| **ECEC organisations** | "child care" OR childcare OR "early childhood" OR kindergarten OR preschool OR pre-school OR pre-primary OR pre-primary OR nursery OR "day care" OR daycare | World-wide differences in the nomenclature of organisations providing education and care services to young children prior to compulsory schooling (Plumb et al. 2013) |
| **Barriers** | barrier OR constrain* OR concern OR limitation OR problem OR obstacle OR hurdle OR risk | Synonyms for barrier |
| **IT** | technolog* OR comput* OR PC OR mobile OR laptop OR notebook OR "interactive whiteboard" OR "interactive white board" OR "electronic whiteboard" OR "smart board" OR smartboard OR "touch screen tablet" OR "tablet computer" OR "tablet computing" OR "tablet personal computer" OR "tablet PC" OR iPad OR Android OR software OR "information system" | Capturing barriers to a range of IT in use within ECEC organisations |
| **ECEC organisation staff** | educator OR teacher OR administrator OR manager OR headmaster OR principal AND NOT (pre-service OR preservice) | Roles of employees in ECEC organisations, excluding students studying to be early childhood educators (known as 'pre-service' educators) |

*Table 1. Literature search criteria and justifications*

Along with the specified keywords, inclusion criterion was set to articles with full-text, in English (the primary author's native language) and from peer-reviewed journals. Additionally, due to a preliminary search of the Scopus database that resulted in a large number of irrelevant articles, an exclusion criterion was added to exclude articles with a subject area of 'medicine'.

Stage *two* involved running search queries combining the keywords above with the inclusion criteria on the selected journal databases. From the searches a total of 625 articles (EBSCO 247, Scopus 235, Web of Science™ 112, EdITLib 31) were identified.

Stage *three* involved the process of filtering the articles. Firstly duplicates of articles that appeared in the results from multiple databases were excluded; then title, abstract and full-text analysis was performed to determine each article's suitability for inclusion in the review. As recommended by Webster and Watson (2002) a backward then forward analysis was performed to enrich the quality of the articles for inclusion (Wolfswinkel et al. 2013): the references in each article were reviewed to determine any articles suitable for inclusion in the review (backward analysis); then the citations of each article were reviewed to determine if they were also suitable for inclusion (forward analysis). This process resulted in a final set of 19 articles for inclusion in the review. The articles reviewed, while not proclaimed to be a definitive list, are representative of the existing body of knowledge on barriers to IT integration within ECEC organisations.

## 3　Results

A quality literature review is one that is based upon a concept-centric approach rather than an author-centric or chronological approach (Webster and Watson 2002). Although many of the articles categorised barriers according to the first-order (obstacles extrinsic to educators) and second-order (internal to educators such as beliefs and attitudes) categories posited by Ertmer (1999), Nikolopoulou and Gialamas (2013 p.3) note that there is not "a single accepted classification of barriers" and as such we decided not to constrict our categorisation of barriers in this way. Instead, the constant comparative method (Lincoln and Guba 1985) was utilised in order to derive barrier categories to structure the results. Each study was reviewed to determine the barriers identified, which were then grouped into a number of tentative categories. Each time a reviewed article identified a barrier this was compared to the existing categories to determine its suitability for inclusion or as a new stand-alone category. This continued until the barrier categories were saturated and no new categories emerged.

### 3.1　Characteristics of the Literature

The literature can be classified as one of two types: studies which have specifically investigated barriers to the integration of IT in ECEC organisations ($n$ = 6); and studies on IT in ECEC organisations that happen to mention barriers ($n$ = 13) as part of their findings. The studies were a mix of purely statistical quantitative studies that utilised closed-question survey instruments; purely qualitative studies utilising interviews and observations for data collection; and several studies that used a mix of both quantitative and qualitative data collection and analysis techniques. Two articles were descriptive in nature and lacked true empirical findings, but were included due to their ability to contribute to the small number of suitable articles. As previously noted in determining the keywords for the database searches, we found that a number of different terms were utilised to describe both employees and ECEC organisations. In regards to employees, we were able to categorise the different terms as referring to either 'educators' or 'managers'. Variations in terminology were also noted in referring to what the educators were 'doing' with the IT in relation to their work practices; the majority of studies referred to the 'integration' or 'use' of IT by educators and subsequently we use the terminology of 'integration' in this review. The barriers identified in the following section are presented in descending frequency of their occurrence in the literature.

### 3.2　Barriers Identified in the Literature

#### 3.2.1　Educator Beliefs and Attitudes

The existing literature identifies educator beliefs and attitudes as a major barrier to IT integration and this was one of the most frequently cited barriers in the literature. Educators held fears that the integration of IT would have a negative impact on children (Li 2006; Lindahl and Folkesson 2012a; Tsitouridou and Vryzas 2004; Wood et al. 2008). Educators spoke of concerns that computers would be an "alienating force…reducing opportunities for social interactions [with a] risk of misuse or abuse" (Li 2006 p.479). Similarly, educators indicated that perceived adverse effects on children resulting from the integration of computers presented a barrier to IT integration (Tsitouridou and Vryzas 2004). These concerns included computers hindering socialisation, fostering individualism, and that children might become addicted to using the computers. Although some educators expressed outright resistance to the integration of IT (Fenty and McKendry Anderson 2014) others expressed confusion and were unsure about whether IT belongs in an ECEC organisation (Ihmeideh 2009; Joshi et al. 2010).

Several studies referred to educators making 'knowledge claims' (Lindahl and Folkesson 2012a) to justify their lack of IT integration or negative attitude towards it. These claims centred on the social, cognitive and motor skills of young children, with educators describing children as "not mature enough" to use computers (Ihmeideh 2010). Wood et al. (2008) identified the age of the target population as a perceived barrier, resulting in support for using computers only with older children, not with the toddler or pre-school aged children. It does appear however that that the limited motor skills of children only presents a barrier to the integration of particular IT devices such as computers; Blackwell's (2013) study of tablet devices noted that "a frequent comment by teachers was the ease of using the iPad, especially for young students who may not have developed the motor skills necessary for using a computer mouse" (p.240).

Other studies reported 'tradition claims' (Lindahl and Folkesson 2012a) and early childhood norms as barriers to IT integration (Blackwell 2013; Li 2006; Lindahl and Folkesson 2012a; Lindahl and Folkesson 2012b; Nikleia and Despo 2005; Parette et al. 2013). Many early childhood educators held beliefs associated with the 'tradition' of early childhood, where the introduction of IT is viewed as "an intrusion and a threat to traditional practice" such as free play (Lindahl and Folkesson 2012a p.1732). The pre-existing teaching philosophies, existing beliefs and practices of educators are important in influencing IT integration (Blackwell 2013); some educators preferred the traditional ways of early childhood practices to new ways that included IT (Parette et al. 2013), or found that the shift from traditional ways of teaching to those involving IT demanded too much effort, time and commitment (Nikleia and Despo 2005), resulting in those educators who are more familiar with traditional teaching styles displaying resistance to IT (Li 2006).

### 3.2.2 Lack of Knowledge and Skills

Studies identified the lack of specific IT knowledge and skills, and IT-supported pedagogical knowledge and skills, as significant barriers to IT integration. As Judge et al. (2004 in Edwards 2005 p.4) argues, "unless early childhood educators have an appropriate understanding of how the technology works, they will be unable to effectively integrate the computer into the learning environment provided for young children".

Educators require specific IT knowledge and skills in order to be able to operate the IT devices and a number of studies (Edwards 2005; Fenty and McKendry Anderson 2014; Ihmeideh 2010; Leung 2003; Li 2006; Nikleia and Despo 2005; Parette et al. 2013; Plowman and Stephen 2005; Tsitouridou and Vryzas 2004; Wood et al. 2008) reported that educators had a lack of such knowledge and skills. Ihmeideh (2010) and Li (2006) found that few early childhood educators are considered 'IT literate', with 40% or less of educators in the investigated Hong Kong kindergartens of Li's study regarded as 'IT literate'. Wood et al. (2008) reported that a lack of specific IT knowledge and skills were a factor which influenced the ability of the educators to "maintain computer equipment and upgrade software [which] could prevent them from using computers" (p.217). This is an important consideration if the ECEC organisation is experiencing the 'lack of technical support' barrier previously described.

Educators all need to have IT-supported pedagogical knowledge and skills; in other words they need to know how to integrate IT in a way that it is developmentally appropriate for an early childhood setting. In Ihmeideh's 2010 study "more than one half of the teachers interviewed mentioned that they do not know how to guide the young children to learn with computers, as they do not have the knowledge of how to make computers valuable tools for young children" (p.71). She concluded that not employing IT in work practices could be due to "the lack of teachers' knowledge about the important role that computer technology plays in developing children's literacy skills" (p.74). Similarly, Tsitouridou and Vryzas (2004) reported that educators "did not understand the possible ways of using computers" and "did not know how computers could be used in the educational process" (p.35).

The lack of both specific IT and IT-supported pedagogical knowledge and skills could be related to the 'lack of training' barrier which will be described in a section below, as Fenty and McKendry Anderson (2014) reported that "the majority of participants in this study described not feeling adequately prepared by their respective teacher education programs to incorporate technology in the classroom" (p.121). Ihmeideh (2009) stated that "seventeen out of thirty pre-school teachers interviewed (56.6%) indicated that they did not attend any in-service training programmes on the use of technology for children…[and] have little knowledge about the use of technology for children in the early years" (p.333). It is however interesting to note that Nikolopoulou and Gialamas (2013) found that "A-level training", a type of technical training for educators, "was not significantly linked to any barrier-factor" (p.13).

### 3.2.3 Lack of Equipment/Resources

A lack of IT equipment/resources was another often-reported barrier in the literature. Studies indicated that a lack of IT included insufficient hardware such as laptops, notebooks, computers, scanners, cameras and projectors, and a lack of Internet access (Fenty and McKendry Anderson 2014; Ihmeideh 2009, 2010; Joshi et al. 2010; Leung 2003; Liu and Pange 2014; Nikleia and Despo 2005; Nikolopoulou and Gialamas 2013; Wood et al. 2008). Several studies (Ihmeideh 2010; Nikleia and Despo 2005; Leung 2003) indicated a complete absence of IT in participants' organisations, while others (Wood et al. 2008; Fenty and McKendry Anderson 2014) described educator difficulties in accessing IT due to an insufficient amount of equipment. Without adequate IT equipment/resources, there is little opportunity for educators to integrate IT into their work practices. An interesting finding was the apparent emergence of a 'digital divide' for educators working with middle-income children; Blackwell et al.'s (2013) U.S. study found that educators working with middle income children have less access to IT compared to educators of lower-income children, possibly due to "technology funding initiatives targeted at lower-income students, such that the policies miss children in the middle income who also do not have equal access to technology compared to higher-income students" (p.317).

### 3.2.4 Lack of Training

Without adequate training, early childhood educators are unable to develop the confidence, skills and knowledge required to successfully integrate IT into their work practices, and a lack of training was a common barrier identified by many studies (Blackwell et al. 2013; Fenty and McKendry Anderson 2014; Ihmeideh 2009; Li 2006; Nikleia and Despo 2005; Parette et al. 2013; Plowman and Stephen 2005; Wood et al. 2008). Although the majority of studies did not indicate why training was not available to early childhood educators, participants in Ihmeideh's (2009) study of Jordanian kindergartens identified that training courses were expensive and kindergartens did not have adequate funding available to send educators on such courses (p.337). Funding as a barrier to IT integration in ECEC organisations will be discussed separately in a subsequent section. Not only was an insufficient amount of training identified as a barrier but Fenty and McKendry Anderson (2014), Plowman and Stephen (2005) and Wood et al. (2008) additionally identified issues with the quality or content of training when it was available. A lack of training for educators on both how to use the IT devices themselves and how to integrate them into their work practices in appropriate ways stands as a potential barrier to effective IT integration.

### 3.2.5 Classroom Condition Constraints

In ECEC organisations where educators were responsible for large numbers of children in one room at a time, studies reported that these large class sizes were a barrier to IT integration (Nikolopoulou and Gialamas 2013; Ihmeideh 2010; Tsitouridou and Vryzas 2004). With large numbers of children in a room there were challenges in regards to managing and supervising the children's access to IT such as computers, which educators reported as making the use of the computers impractical (Wood et al. 2008 p.223). Educators also felt constrained in their use of IT by the perceived negative impact that waiting had on children's behaviour, with reports of children becoming aggressive while waiting to take turns on an interactive whiteboard (Fenty and McKendry Anderson 2014).

The physical location of IT was also noted as a barrier to their integration into educator work practices (Fenty and McKendry Anderson 2014; Edwards 2005). Educators reported difficulties with accessing digital cameras that were not kept in the classroom but rather in the main office for shared access (Fenty and McKendry Anderson 2014). Educators in Edwards' (2005) study reported that the location of computers in the organisation's offices limited their integration of IT, as they had to ensure children were supervised while using the computer in the offices. Blackwell's (2013) study of tablet devices included a participant who reported constraints on using iPads in their organisation as a result of having to manage a class set of 42 iPads; this included the frustration of managing updates on each iPad individually and being unable to use them when they ran out of battery power.

### 3.2.6 Educator Lack of Confidence

Educators' lack of confidence in using IT creates a significant barrier to IT integration. Blackwell et al. (2014), Fenty and McKendry Anderson (2014), Joshi et al. (2010), Li (2006), Nikolopoulou and Gialamas (2013), Plowman and Stephen (2005) and Tsitouridou and Vryzas (2004) all reported educators who had very low levels of confidence in their abilities to integrate IT, with descriptive terms like 'discomfort' and 'intimidated' often used by the educators when describing how they felt about IT. Blackwell et al. (2014) suggests that educator confidence plays a large role in shaping attitudes towards the use of IT (a barrier to IT integration described above), and Nikolopoulou and Gialamas (2013) identified confidence as a moderating influence on the barriers of lack of financial support, technical support and class conditions.

### 3.2.7 Lack of Appropriate Educational Software

Access to appropriate educational software to run on the IT hardware was frequently reported as a barrier to IT integration (Blackwell 2013; Edwards 2005; Ihmeideh 2009; Li 2006; Liu and Pange 2014; Nikolopoulou and Gialamas 2013; Plowman and Stephen 2005). The significance of this barrier was conveyed by Ihmeideh (2009) who stated that "most pre-school teachers believed that the most serious barrier facing the use of technology in pre-school settings is related to software" (p.332). Nikolopoulou and Gialamas (2013) concurred, finding that "Greek teachers also perceived as a major barrier the lack of appropriate/good educational software" (p.12), Similarly, in Liu and Pange's (2014) study, a lack of appropriate content on IT devices was the top barrier perceived by educators. It was not just the lack of access to appropriate software that the studies identified, but also issues educators had in determining what constituted appropriate software. For example, Plowman and Stephen (2005) identified that educators required "help with identifying appropriate software" (p.155) and a participant in Edwards' (2005) study described how her ability to select software for use was limited by her knowledge of the computer's specifications. Although most studies identified software-related barriers in regards to computers, Blackwell (2013) also found that "teachers reported difficulties in finding and selecting iPad apps for their students to use, especially for teachers with special [needs] student populations" (p.242).

### 3.2.8 Lack of Support

The majority of studies identified a lack of support as a significant barrier to IT integration, with Blackwell et al. (2013) describing support as "critical to technology integration in the early childhood classroom" (p.87). Where authors were specific about support-related barriers, they were identified as belonging to one of two different categories: (i) technical support (Li 2006; Nikolopoulou and Gialamas 2013; Plowman and Stephen 2005; Wood et al. 2008), and (ii) administrative/stakeholder support (Fenty and McKendry Anderson 2014; Li 2006; Liu and Pange 2014; Nikleia and Despo 2005; Nikolopoulou and Gialamas 2013). A lack of technical support for educators was identified as the second-highest ranked barrier in Nikolopoulou and Gialamas' (2013) study, who additionally noted that educators with more years of computer experience and more confidence in IT perceived the lack of support as only a minor barrier. A contrasting finding by Liu and Pange (2014) was that educators who were using IT in daily life (and were therefore more experienced with IT) were more likely to perceive a lack of support as a barrier (p.10). Without adequate technical support educators faced technical challenges such as computer breakdowns, which were perceived to be "frustrating, and in some cases prohibitive" (Wood et al. 2008 p.224).

Inadequate support from stakeholders such as administrators and parents was also identified as a barrier. If administrators or managers did not support the integration of IT, preferring instead a traditional early childhood setting without IT (as identified in Nikleia and Despo's 2005 study), then this becomes a significant barrier to the integration of IT in an ECEC organisation. As parents are considered to be influential stakeholders in the environment of an educational organisation (Burden et al. 2012 cited in Clark and Luckin 2013), it is unsurprising to find several studies which identified a lack of support from parents as a barrier to IT integration, particularly from those parents who are considered to be not 'IT literate' (Li 2006); although in Liu and Pange's (2014) study this lack of parental support was statistically one of the least prominent barriers.

### 3.2.9 IT Technical Problems

Outdated, incompatible and unreliable IT can cause significant constraints on educators' integration of IT (Blackwell 2013; Edwards 2005; Fenty and Anderson 2014; Ihmeideh 2009; Li 2006; Nikolopoulou and Gialamas 2013). Educators in Edwards' (2005) study reported frustrations and limitations caused by computers "freezing", or "working too slowly", with one educator reporting that a computer that "tended to shut down without obvious reason" (p.5) and concluded that it was "simply considered easier to avoid using the computer at all" (p.5). Similar frustrations were expressed in Blackwell's (2013) study of integrating tablet devices into an ECEC organisation, where educators frequently listed 'unreliability' as a major drawback to using iPad devices, specifically identifying the constraints of apps not working and an inconsistent Internet connection as barriers to their effective integration (p.242). The age of the IT that was available to educators was also identified as a barrier to integration (Fenty and McKendry Anderson 2014; Ihmeideh 2009; Nikolopoulou and Gialamas 2013). In these studies participants made specific comments that although they had access to IT, the IT was old and outdated which caused barriers to its integration; in Fenty and McKendry Anderson's (2014) study the IT was "just too old" for the newer games they wanted to install and use (p.123).

### 3.2.10 Lack of Funding

Funding is perhaps implicit in underpinning many barriers identified such as 'lack of access to equipment/resources' and 'lack of training', yet a number of studies (Ihmeideh 2009; Li 2006; Nikolopoulou and Gialamas 2013; Parette et al. 2013; Plowman and Stephen 2005; Wood et al. 2008) explicitly identified funding or budget limitations as a barrier to IT integration in ECEC organisations, with a lack of funding found to be the "leading perceived barrier" by educators in the study by Nikolopoulou and Gialamas (2013 p.12). Ihmeideh (2009) reported that 13 out of 15 principals indicated that "technology is not widely used in pre-school settings because of [a] lack of funding" (p.334) and stated that pre-schools in their country are run by the private sector where financial support from the government is not available. Li's (2006) study also involved principals and had one principal reporting (and others concurring) that "most of the kindergartens are privately owned…and their major income is tuition fee…many early childhood settings including mine have experienced substantial difficulties in raising funds to enable construction of ICT infrastructure and the integration" (p.478). Ihmeideh also reported in her 2010 study that "training programs are costly and that is why our kindergarten administration does not want us to attend and participate in these programs" (p.71). These findings highlight a link between 'lack of funding' and the 'nature of the early childhood education sector' barrier described later in this section.

### 3.2.11 Physical Environment Constraints

Physical environmental conditions of ECEC organisations can result in barriers to IT integration (Edwards 2005; Ihmeideh 2009; Ihmeideh 2010; Li 2006; Tsitouridou and Vryzas 2004; Wood et al. 2008). The size of rooms and the available space in them were reported as problematic (Ihmeideh 2009; Wood et al. 2008), in addition to the limited availability of electrical outlets (Edwards 2005; Wood et al. 2008). Ihmeideh's studies (2009, 2010) conducted in Jordan reported that the majority of kindergartens in the country were initially houses and not specifically built as kindergartens, resulting in physical locations that were not adequately equipped for the use of technology. A similar finding occurred in Tsitouridou and Vryzas's (2004) study in Greece where the buildings were deemed not suitable and material-technical infrastructure was inadequate. However, such physical barriers were not reported in studies from other countries.

### 3.2.12 Lack of Time

Educators require time to learn to use new IT, find or develop resources and generally prepare for IT utilisation in an ECEC organisation. Not having adequate time to undertake these activities was reported as a barrier to IT integration by Fenty and McKendry Anderson (2014), Ihmeideh (2009, 2010), Li (2006) and Wood et al. (2008). Early childhood educators are not just responsible for providing educational activities for children; they are also required to undertake care duties such as toileting and assisting at meal times, supervising play-time, liaising with parents, and undertaking administrative duties. In the literature there was an acknowledgement that the daily schedule of an educator is busy and finding time to prepare for and integrate IT into this daily schedule was reported as difficult (Wood et al. 2008). An overloaded curriculum (Li 2006) was also reported as a reason why educators were short on time to devote to successful IT integration.

### 3.2.13 Early Childhood Curriculum and Guidelines

Although countries such as Australia, New Zealand, the U.K., Portugal, Sweden, and Denmark now include specific mentions of IT in their early childhood curricula, a number of studies reported a lack of clear or concrete guidelines on how to integrate IT into early childhood educator practices. Although previous research has shown "how a number of international, governmental institutions, as well as private organisations, have interest in and produce recommendations for, the use of ICT in preschool practice" (Lindahl and Folkesson 2012a p.1730), Ljung-Djärf (2008) believes that evidence-based guidelines for IT in early childhood education are still limited. A lack of pedagogical models was identified as a leading barrier by educators (Liu and Pange 2014) and in the literature several accounts exist of early childhood educators reporting that a lack of guidelines constrained them in being able to translate their support for IT into practical and innovative opportunities to integrate IT into their curriculum (Blackwell 2013; Liu and Pange 2014; Wood et al. 2008).

### 3.2.14 Nature of the Early Childhood Educational Sector

The nature of the early childhood educational sector and the settings in which ECEC organisations exist may also present potential barriers to IT integration. Plowman and Stephen (2005) and Parette et al. (2013) mention the distinct culture that exists around the sector compared with the school educational sector, with Parette et al. suggesting that "early childhood education programs are distinct cultural groups with varying values, behaviours, and characteristics. These programs mirror the communities within which they reside, and it is not uncommon to encounter resistance to technology use…if the community has values that have led to a recognized tradition of delivering the curriculum in ways that are not supported by technology" (p.4). As ECEC organisations can vary in their type, for example for/not-for profit, school/non-school based, public/private, government/non-government funded, this could result in specific barriers not found in other educational sectors. As Wood et al. (2008) state, "supporting computer technology in the early childhood education environment may be a particular challenge because [in the U.S. at least] these programs are not government-funded, networked, or organised through a central administration unit, hence isolating each centre and increasing the pressures on individual early childhood educators" (p.224).

### 3.3 Summary

The identified barriers from the literature and the papers in which these barriers are mentioned are summarised in Table 2 below.

| Barriers | Blackwell (2013) | Blackwell et al. (2013) | Blackwell et al. (2014) | Edwards (2005) | Fenty and McKendry Anderson (2014) | Ihmeideh (2009) | Ihmeideh (2010) | Joshi et al. (2010) | Leung (2003) | Li (2006) | Lindahl and Folkesson (2012a) | Lindahl and Folkesson (2012b) | Liu and Pange (2014) | Nikleia and Despo (2005) | Nikolopoulou and Gialamas (2013) | Parette et al. (2013) | Plowman and Stephen (2005) | Tsitouridou and Vryzas (2004) | Wood et al. (2008) |
|---|---|---|---|---|---|---|---|---|---|---|---|---|---|---|---|---|---|---|---|
| Educator Beliefs and Attitudes | ✓ | | | | ✓ | ✓ | ✓ | ✓ | | ✓ | ✓ | ✓ | | ✓ | | ✓ | | ✓ | ✓ |
| Lack of Knowledge and Skills | | | | ✓ | ✓ | ✓ | ✓ | | ✓ | ✓ | | | | ✓ | ✓ | ✓ | ✓ | ✓ | ✓ |
| Lack of Equipment/Resources | | ✓ | | | ✓ | ✓ | ✓ | ✓ | ✓ | | | | ✓ | ✓ | ✓ | | | | ✓ |
| Lack of Training | | ✓ | | | ✓ | ✓ | | | | ✓ | | | | ✓ | | ✓ | ✓ | | ✓ |
| Classroom Condition Constraints | ✓ | | | ✓ | ✓ | | ✓ | | | | | | | | ✓ | | | ✓ | ✓ |
| Educator Lack of Confidence | | | ✓ | | ✓ | | | ✓ | | ✓ | | | | | ✓ | | ✓ | ✓ | |
| Lack of Appropriate Educational Software | ✓ | | ✓ | | ✓ | | | | | ✓ | | ✓ | | ✓ | | ✓ | | | |
| Lack of Support | | | | | | ✓ | | | | ✓ | | | ✓ | ✓ | | | ✓ | | ✓ |
| IT Technical Problems | ✓ | | | ✓ | ✓ | ✓ | | | | ✓ | | | | | ✓ | | | | |
| Lack of Funding | | | | | | ✓ | | | | ✓ | | | | | ✓ | ✓ | ✓ | | ✓ |
| Physical Environment Constraints | | | | ✓ | | ✓ | ✓ | | | ✓ | | | | | | | | ✓ | ✓ |
| Lack of Time | | | | | ✓ | ✓ | ✓ | | | ✓ | | | | | | | | | ✓ |
| Early Childhood Curriculum and Guidelines | ✓ | | | | | | | | | | | | | ✓ | | | | | ✓ |
| Nature of the Early Childhood Educational Sector | | | | | | | | | | | | | | | | ✓ | ✓ | | ✓ |

*Table 2. Summary of identified barriers from the literature*

## 4 Discussion: Knowledge Gaps and Existing Literature Analysis

### 4.1 Perceived vs. Actual Experience of Barriers to IT Integration

Many studies on barriers to IT integration in ECEC organisations refer to 'perceived' barriers by educators. However, a barrier that is 'perceived' does not necessarily exist in reality and may not have actually been experienced by the educator. Despite the barriers 'perceived' by educators, Nikolopoulou and Gialamas (2013) found that 67.2% of participants reported that they use computers in class; the 'perceived' barriers did not necessarily stop the IT integration. Blackwell et al.'s (2013) quantitative findings indicated that no barrier factors predicted IT use and they suggest that although educators may feel limited by certain barriers, this is not consistent with practice; the educators do not necessarily use the IT less, instead they "may not be using it in ways and to the extent they desire or feel the technology affords" (p.317).

### 4.2　Educators' Ways of Dealing With Barriers to IT Integration

The majority of studies did not report on how the educators dealt with the barriers and how their work practices were impacted as a result. Most studies made brief general suggestions and recommendations on how the barriers may be addressed; for example Liu and Pange (2014) make recommendations such as "sufficient ICT-related equipment and curriculum resource/content should be provided to these teachers" and "effective and quality pre-service and in-service training is needed" (p.12) but did not expand any further. Three exceptions were the studies by Li (2006) and Leung (2003) who both reported that educators worked around the barrier of lack of appropriate software by using Microsoft PowerPoint to create self-made software activities for the children; and Blackwell's (2013) study where an educator described how they dealt with the situation of lacking enough devices for each student by using the tablets to develop the students' skill in sharing devices (p.250).

### 4.3　Focus on Educators

The majority of articles only involved participants who were educators, which reveals barriers to IT integration from their perspectives but does not assist in identifying additional barriers that may be identified by other stakeholders such as managers, leaders, or parents. Only two studies drew participants from roles other than that of 'educator': Ihmeideh (2009) interviewed both educators and principals of kindergartens; and Li's (2006) study involved principals of ECEC organisations.

### 4.4　Form of IT

Apart from Blackwell et al. (2013, 2014), Fenty and McKendry Anderson (2014) and Blackwell (2013), the studies reviewed identified barriers in regards to the use of desktop personal computers (PCs). Plowman and Stephen (2003) suggest that most of the literature on IT integration in educational organisations refers to IT as synonymous with desktop PCs and that this is problematic as it results in the identification of barriers that may be specific to desktop PCs and that such barriers have "less currency" (p.153) with other forms of IT such as tablet devices. As Plowman and Stephen (2003) suggest, many of the concerns about children's use of IT are "based on a concept of technology that is now out of date" (p.157). Previous studies of desktop PC integration (Wood et al. 2008) have identified that "younger children did not have the fine motor control necessary to use computers effectively" (p.218) which presented a barrier to their integration. However the educators in Blackwell's (2013) study of tablet devices did not report this same barrier in regards to the children's fine motor control; instead they reflected that when using iPads with young children in their work practices, the children "may not have developed the motor skills necessary for using a computer mouse" (p.240), and that there was a distinct ease of use arising from the touch screen interface of the iPads. There are also possibilities for barriers specific to the newer forms of IT such as mobile tablet devices. These possible barriers, such as privacy issues, have not been previously identified in the existing body of literature but have recently been recognised by Plumb and Kautz (2014a). Such newly identified barriers are related to both the physical form of the IT and the work practices of the early childhood educators in which the IT device is involved.

### 4.5　Type of ECEC Organisation

Only two barrier-specific studies drew participants from a range of different types of ECEC organisations. Blackwell et al. (2013) identified participants in their survey as educators working in a range of different ECEC organisations that exist in the U.S. such as non-school care centres (for-profit or non-profit), school-based care centres (public and private), and home-based childcare (p.312); while Wood et al. (2008) specifically targeted particular types of ECEC organisations to obtain "a diverse array of possible early childhood settings including large and small, public and private daycare centres (that provided both full-time and/or part-time care for children) and university laboratory preschool centres" (p.213). The significance of investigating different types of ECEC organisations was revealed by Blackwell et al. (2013) who presented findings highlighting differences between the types of ECEC organisations and the extent to which educators had access to technology, positing that the suggestion of 'universal IT access' is not necessarily accurate. Ihmeideh (2009; 2010) supports the importance of understanding how barriers differ in different types of ECEC organisations as she found a link between the 'lack of funding' barrier and the educators' different beliefs about the use of computers in teaching in publicly funded and private kindergartens.

### 4.6 Limitations of Closed-Question Surveys as Instruments for Identifying Barriers

Studies utilising closed-question survey items are only able to provide information on the specific barriers that are questioned for. The majority of barrier-specific studies (cf. Blackwell et al. 2013, 2014; Liu and Pange 2014; Nikolopoulou and Gialamas 2013) utilised quantitative closed-question surveys that statistically analysed pre-defined barriers identified from previous studies, often from studies of IT integration in schools. An exception was the study by Ihmeideh (2009) who utilised qualitative semi-structured interviews to collect data that was then analysed to identify barriers. Notably Nikolopoulou and Gialamas (2013) used Ihmeideh's study in formulating the barrier questions in their survey instrument, acknowledging the limitations of utilising closed-question surveys and "the use of a quantitative inquiry only" and suggested open-ended questions to "help understand the importance of barriers when it comes to integrating technology in the classroom" (p.14). Liu and Pange (2014) also noted in their limitations that only a "selection" of first-order and second-order barrier variables were utilised and suggest both qualitative and quantitative methods in future studies "in order to obtain a better understanding of the barriers" (p.13).

### 4.7 Similarities of Identified Barriers with School and University Educational Organisations

Many authors acknowledge that the barriers identified by educators in ECEC organisations are similar to those identified in school and university educational organisations (c.f. Blackwell et al. 2013; Fenty and McKendry Anderson 2014; Liu and Pange 2014; Nikolopoulou and Gialamas 2013), particularly those barriers related to "not having an adequate technological and pedagogical knowledge base, limited skill set, and inadequate access to up-to-date technology and training" (Fenty and McKendry Anderson 2014 p.125). Although this could be true, it is suggested that because many of the studies reviewed utilised closed-question surveys that drew from the school-based literature, there would be an associated bias towards such findings. This makes it significant for future studies into barriers to IT integration in ECEC organisations to utilise open-ended questions and other qualitative data collection techniques to provide opportunities for ECEC organisational educators and management to identify any barriers they have experienced, without being constrained to previously identified barriers, particularly those from other educational organisations.

## 5   A Categorisation of Barriers and their Relationships

Many of the studies reviewed made reference to the work of Ertmer (1999) who developed a classification of what are termed 'first-order' and 'second-order' barriers to educators within schools who are integrating IT into their work practices. First-order barriers are considered those "obstacles that are extrinsic to teachers" (p.50) and include resources such as equipment, time and training that are missing or inadequate within an educator's environment; second-order barriers are those that are "typically rooted in teacher's underlying beliefs about teaching and learning" (p.51).

In their study of the appropriation of IT in an ECEC organisation as "the way that users evaluate and adopt, adapt and integrate a technology into their everyday practices" (Mendoza et al. 2010 p.6), Plumb and Kautz (2014b) adapted and utilised a tri-perspective framework which was originally proposed by Slappendel (1996) to categorise the literature on innovation in organisation. We propose that this framework can be utilised in this literature review in order to categorise the barriers in an innovative manner and to highlight a further gap in the literature for future research into a more process-oriented understanding of how barriers interact and are influential at different stages of the IT integration process. The *individualist* perspective of the framework categorises barriers related to actions and personality traits, similar to the 'second-order' classification of barriers by Ertmer (1999). The barriers identified from the literature and categorised into this perspective are: 'educator beliefs and attitudes', 'lack of knowledge and skills', and 'educator lack of confidence' and 'lack of time'. The *structuralist* perspective categorises barriers that are related to organisational characteristics and those that result from the environment that surrounds an ECEC organisation. From the existing literature this includes: 'lack of equipment/resources', 'lack of support', 'lack of training', 'lack of time', 'physical environment constraints', 'classroom condition constraints', 'IT technical problems', 'lack of appropriate educational software', 'lack of funding', 'nature of early childhood education sector' and 'early childhood curriculum and guidelines'. The *interactive process* perspective is of particular interest as it views the integration of IT as a dynamic, continuous process of change, where individualist and structuralist elements interact and shape the process of IT integration. It can be used to understand how and why barriers are present and influential at certain stages of the integration process, in addition to exploring the relationships between barriers during the integration process. The temporality of barriers is worthy of investigation as "different barriers are likely to appear at different points in the integration process" (Ertmer 1999 p.53). Additionally barriers may never be eliminated completely, as Ertmer (1999) suggests instead that they will "continue to ebb and flow throughout the evolutionary integration process (Becker 1993). At some points, first-order barriers will be at the forefront; at other times, second-order barriers will present the more critical challenges" (p.52).

This literature review has revealed no articles exist that examine barriers to IT integration from a process perspective, although Wood et al. (2005 in Nikolopoulou and Gialamas 2013) call for evaluating "the impact of potential barriers over time" (p.14) and Nikolopoulou and Gialamas themselves (2013) suggest further research to understand "how some previously reported barriers have changed" (p.14). An interactive process perspective would also permit a better understanding of the relationship between individualist and structuralist barriers during the process of IT integration. Ertmer (1999) suggests that the relationships between barriers to IT integration appear "much more complex than initially proposed" (p.52), a finding concurred by Liu and Pange (2014) who suggest that in the educational organisation literature, the relationship between barriers to IT integration was complex and barriers were closely interrelated. A preliminary analysis of the identified barriers suggests that there are indeed close relationships between several barriers. For example: a lack of funding influences the lack of equipment/resources, lack of training and lack of support; a lack of time influences lack of training; and a lack of training influences a lack of knowledge and skills, which in turn influences educator lack of confidence. However there was only one article in the body of literature that attempted to understand the relationships between barriers. Blackwell et al.'s (2014) study investigated the "associations between extrinsic and intrinsic barriers to technology use for teachers of very young children" (p.88) but was limited by only examining the relationship between the identified barriers of support, technology policy, student socioeconomic status, teaching experience, confidence and attitude and how they influenced the variance of IT use in a static snapshot of users' perceived barriers. Table 3 below presents a tri-perspective framework for understanding barriers to the integration of IT in ECEC organisations based on the existing literature.

| Perspective | Focus | Concept | Sources |
|---|---|---|---|
| **Individualist** | Barriers identified related to an individual's internal characteristics and traits | Educator beliefs and attitudes, lack of knowledge and skills, educator lack of confidence | Blackwell et al. 2014; Fenty and McKendry Anderson 2014; Edwards 2005; Ihmeideh 2009, 2010; Joshi et al. 2010; Leung 2003; Li 2006; Lindahl and Folkesson 2012a; Nikleia and Despo 2005; Nikolopoulou and Gialamas 2013; Parette et al. 2013; Plowman and Stephen 2005; Tsitouridou and Vryzas 2004; Wood et al. 2008 |
| **Structuralist** | Barriers identified related to organisational characteristics and elements of the environment surrounding an ECEC organisation | Lack of equipment/resources, lack of support, lack of training, lack of time, physical environment constraints, classroom condition constraints, IT technical problems, lack of appropriate educational software, lack of funding, nature of early childhood education sector, early childhood curriculum and guidelines | Blackwell et al. 2013; Edwards 2005; Fenty and McKendry Anderson 2014; Ihmeideh 2009, 2010; Joshi et al. 2010; Leung 2003; Li 2006; Liu and Pange 2014; Nikleia and Despo 2005; Nikolopoulou and Gialamas 2013; Parette et al. 2013; Plowman and Stephen 2005; Wood et al. 2008 |
| **Interactive Process** | Barriers are related to the integration of IT into work practices of ECEC organisational employees. By viewing IT integration as a complex and dynamic process, an understanding of the nature, temporality, influence of and relationship between barriers during this process can be developed. | | |

*Table 3: A tri-perspective framework for understanding barriers to the integration of IT in ECEC organisations*

# 6 Conclusion

This literature review has revealed that there is a scarcity of empirical studies examining barriers to IT integration within ECEC organisations. Many studies identified barriers in common with schools, however Stephen and Plowman (2003) suggest that it is "not possible to extrapolate confidently from these school-based studies" (p.228). Due to the unique features of the early childhood educational sector such as the influence of historical debate on the place of IT in young children's learning and development (Blackwell et al. 2014; Ihmeideh 2009; Wood et al. 2008) and the diverse range of educational and training backgrounds of employees (Plowman and Stephen 2003), there may be barriers which are particular to ECEC organisations. Blackwell et al. (2013) indicate that there are important differences between early childhood educators and school educators and there is a need for future research to "disaggregate findings for these two teacher demographics" (p.318). Other identified gaps included the difference between perceived versus actual experience of barriers and their relation to educator practices, and the limitations of utilising closed-question statistical survey data collection and analysis, which constrains findings. As these gaps in the small body of existing literature have been recognised, a tri-perspective framework is proposed which in conjunction with an open-ended qualitative data collection and methodology can move to address the calls for further research into barriers to IT integration within ECEC organisations. The framework can also provide an avenue for a deeper, more detailed understanding of not only what barriers exist but how they interrelate and shape the IT integration process and the work practices of ECEC organisational employees.

# 7 References

Association for Information Systems. 2015. "Senior Scholars' Basket of Journals" http://aisnet.org/?SeniorScholarBasket Retrieved 2 February 2013.

Blackwell, C. 2013. "Teacher practices with mobile technology integrating tablet computers into the early childhood classroom," *Journal of Education Research* (7:4), pp. 231-255.

Blackwell, C.K., Lauricella, A.R., and Wartella, E. 2014. "Factors influencing digital technology use in early childhood education," *Computers & Education* (77:0), pp. 82-90.

Blackwell, C.K., Lauricella, A.R., Wartella, E., Robb, M., and Schomburg, R. 2013. "Adoption and use of technology in early education: The interplay of extrinsic barriers and teacher attitudes," *Computers & Education* (69:0), pp. 310-319.

British Educational Communications and Technology Agency. 2004. "A Review of the Research Literature on Barriers to the Uptake of ICT by Teachers".

Clark, W., and Luckin, R. 2013. "What the research says: iPads in the Classroom". London Knowledge Lab, Institute of Education, University of London.

Edwards, S. 2005. "Identifying the factors that influence computer use in the early childhood classroom," *Australasian Journal of Educational Technology* (21:2), pp. 192-210.

Elsevier B.V. 2015. "Scopus | Elsevier" http://www.elsevier.com/online-tools/scopus Retrieved 10 February 2015.

Ertmer, P.A. 1999. "Addressing first- and second-order barriers to change: Strategies for technology integration," *Educational Technology Research and Development* (47:4), pp. 47-61.

Fenty, N.S., and McKendry Anderson, E.M. 2014. "Examining Educators' Knowledge, Beliefs, and Practices About Using Technology With Young Children," *Journal of Early Childhood Teacher Education* (35:2), pp. 114-134.

Global U. 2015. "EdITLib Contents" http://www.editlib.org/about/contents/ Retrieved 10 February 2015.

Hew, K., and Brush, T. 2007. "Integrating technology into K-12 teaching and learning: current knowledge gaps and recommendations for future research," *Educational Technology Research and Development* (55:3), pp. 223-252.

Ihmeideh, F.M. 2009. "Barriers to the use of technology in Jordanian pre-school settings," *Technology, Pedagogy and Education* (18:3), pp. 325-341.

Ihmeideh, F.M. 2010. "The role of computer technology in teaching reading and writing: preschool teachers' beliefs and practices," *Journal of Research in Childhood Education* (24:1), pp. 60-79.

Joshi, A., Pan, A., Murakami, M., and Narayanan, S. 2010. "Role of Computers in Educating Young Children: U.S. and Japanese Teachers' Perspectives," *Computers in the Schools* (27:1), pp. 5-19.

Leung, W.M. 2003. "The Shift from a Traditional to a Digital Classroom Hong Kong Kindergartens," *Childhood Education* (80:1), pp. 12-17.

Li, H. 2006. "Integrating Information and Communication Technologies Into the Early Childhood Curriculum: Chinese Principals' Views of the Challenges and Opportunities," *Early Education and Development* (17:3), pp. 467-487.

Lincoln, Y.S., and Guba, E.G. 1985. *Naturalistic inquiry*. Beverly Hills, CA: Sage Publications, Inc.

Lindahl, M.G., and Folkesson, A.-M. 2012a. "Can we let computers change practice? Educators' interpretations of preschool tradition," *Computers in Human Behavior* (28:5), pp. 1728-1737.

Lindahl, M.G., and Folkesson, A.-M. 2012b. "ICT in preschool: friend or foe? The significance of norms in a changing practice," *International Journal of Early Years Education* (20:4), pp. 422-436.

Liu, X., and Pange, J. 2014. "Early childhood teachers' perceived barriers to ICT integration in teaching: a survey study in Mainland China," *Journal of Computers in Education*, pp. 1-15.

Ljung-Djärf, A. 2008. "To Play or Not to Play - That Is the Question: Computer Use Within Three Swedish Preschools," *Early Education and Development* (19:2), pp. 330-339.

Mendoza, A., Carroll, J., and Stern, L. 2010. "Software Appropriation over Time: From Adoption to Stabilization and Beyond," *Australasian Journal of Information Systems* (16:2), pp. 5-23.

Nikleia, E., and Despo, K. 2005. "Computer based early childhood learning," in *Proceedings of EUROCON 2005. The International Conference on Computer as a Tool*, Belgrade, Serbia, November 21-24.

Nikolopoulou, K., and Gialamas, V. 2013. "Barriers to the integration of computers in early childhood settings: Teachers' perceptions," *Education and Information Technologies*, pp. 1-17.

Parette, H.P., Blum, C., and Quesenberry, A.C. 2013. "The role of technology for young children in the 21st century," in *Instructional technology in early childhood,* H.P. Parette and C. Blum (eds.). Brookes Publishing, pp. 1-28.

Plowman, L., and Stephen, C. 2003. "A 'benign addition'? Research on ICT and pre-school children," *Journal of Computer Assisted Learning* (19:2), pp. 149-164.

Plowman, L., and Stephen, C. 2005. "Children, play, and computers in pre-school education," *British Journal of Educational Technology* (36:2), pp. 145-157.

Plumb, M., Kautz, K., and Tootell, H. 2013. "Touch screen technology adoption and utilisation by educators in early childhood educational institutions: A review of the literature," in *Proceedings of 24th Australasian Conference on Information Systems (ACIS 2013)*, Melbourne, Australia, December 4-6.

Plumb, M., and Kautz, K. 2014a. "Reconfiguring Early Childhood Education and Care," in *Information Systems and Global Assemblages. (Re) Configuring Actors, Artefacts, Organizations*, B. Doolin, E. Lamprou, N. Mitev, and L. McLeod (eds), Springer Berlin Heidelberg, pp. 30-47.

Plumb, M., and Kautz, K. 2014b. "Innovation within an Early Childhood Education and Care Organisation: A Tri-Perspective Analysis of the Appropriation of IT," in *Proceedings of 25th Australasian Conference on Information Systems (ACIS 2014)*, Auckland, New Zealand, December 8-10.

Reid, P. 2014. "Categories for barriers to adoption of instructional technologies," *Education and Information Technologies* (19:2), pp. 383-407.

Slappendel, C. 1996. "Perspectives on innovation in organizations", *Organization Studies* (17:1), pp. 107-129.

State Government Victoria (Department of Education and Early Childhood Development). 2013. "National Quality Framework - Supervision" http://www.education.vic.gov.au/ Retrieved 10 February 2015.

Tsitouridou, M., and Vryzas, K. 2004. "The prospect of integrating ICT into the education of young children: the views of Greek early childhood teachers," *European Journal of Teacher Education* (27:1), pp. 29-45.

Webster, J., and Watson, R.T. 2002. "Analyzing the Past to Prepare for the Future: Writing a Literature Review," *MIS Quarterly* (26:2), pp. xiii-xxiii.

Wolfswinkel, J.F., Furtmueller, E., and Wilderom, C.P.M. 2013. "Using grounded theory as a method for rigorously reviewing literature," *European Journal of Information Systems* (22:1), pp. 45-55.

Wood, E., Specht, J., Willoughby, T., and Mueller, J. 2008. "Integrating Computer Technology in Early Childhood Education Environments: Issues Raised by Early Childhood Educators," *Alberta Journal of Educational Research* (54:2), pp. 210-226.